\documentstyle[osa,psfig,multicol]{revtex} 

\begin{document}

\title{Generic features of modulational instability in nonlocal Kerr media.}

\author{John Wyller}
\address{Department of Mathematical Sciences, Agricultural University 
of Norway, P.O.Box 5065, N-1432 \AA s, Norway.}

\author{Wieslaw Krolikowski}
\address{Australian Photonics Cooperative Research Centre, 
Laser Physics Centre, Research School of Physical 
Sciences and Engineering, Australian National
University, Canberra ACT 0200, Australia.}

\author{Ole Bang}
\address{Department of Informatics and Mathematical Modelling,
Technical University of Denmark, DK-2800 Kongens Lyngby, Denmark.}

\author{Jens Juul Rasmussen}
\address{Ris\o\ National Laboratory, Optics and Fluid Dynamics Department, 
OFD - 128, P. O. Box 49, DK-4000 Roskilde, Denmark.}

\maketitle

\begin{abstract}
The modulational instability (MI) of plane waves in nonlocal Kerr media 
is studied for a general, localized, response function. 
It is shown that there always exists a finite number of well-separated MI
gain bands, with each of them characterised by a unique maximal growth rate. 
This is a general property and is demonstrated here for the Gaussian, 
exponential, and rectangular response functions.  
In case of a focusing nonlinearity it is shown that although the nonlocality  
tends to suppress MI, it can never remove it completely, irrespectively of 
the particular shape of the response function.
For a defocusing nonlinearity the stability properties depend sensitively 
on the profile of the response function. 
It is shown that plane waves are always stable for response functions with 
a positive-definite spectrum, such as Gaussians and exponentials. 
On the other hand, response functions whose spectra change sign (e.g., 
rectangular) will lead to  MI in the high wavenumber regime, provided the 
typical length scale of the response function exceeds a certain threshold.
Finally, we address the case of generalized multi-component response functions
consisting of a weighted sum of N response functions with known properties .\\

\end{abstract}

\begin{multicols}{2}

\newlength{\figwidthb}
\setlength{\figwidthb}{0.95\linewidth}

\section{Introduction}

The phenomena of modulational instability (MI) of plane waves has been
identified and studied in various physical systems, such as fluids 
\cite{mi-fluid}, plasma \cite{mi-plasma}, nonlinear optics 
\cite{mi-nlo1,mi-nlo2}, discrete nonlinear systems, such as molecular 
chains \cite{discrete-mi-molchain} and Fermi-resonant interfaces and 
waveguide arrays \cite{discrete-mi-array}, dispersive nonlinear 
directional couplers with the change of refractive index following a 
exponential relaxation law \cite{trillo} etc. 
It has been shown that MI is strongly affected by various
mechanisms present in nonlinear systems, such as higher order dispersive
terms in the case of optical pulses \cite{mi-dispersive}, saturation of the
nonlinearity \cite{mi-saturation}, and coherence properties of optical beams 
\cite{mi-coherence}.

In this work we study the MI of plane waves propagating in a nonlinear Kerr
type medium with a nonlinearity (the refractive index change, in nonlinear
optics) that is a nonlocal function of the incident field. We consider a
phenomenological model 
\begin{equation}
   i\partial_z\psi + \frac{1}{2}\partial_x^2\psi + 
   s\psi\int_{-\infty}^{\infty} R(x^{\prime}-x)|\psi(x^{\prime})|^2
   dx^{\prime} = 0,  
   \label{generalNLS}
\end{equation}
for the wave modulation $\psi $ where $x$ is the transverse spatial
coordinate and $s=1$ $(s=-1)$ corresponds to a focusing (defocusing)
nonlinearity. The evolution coordinate $z$ can be time, as for Bose-Einstein
Condensates (BEC's), or the propagation coordinate, as for optical
beams. We consider only symmetric spatial response functions that are
positive definite and (without loss of generality) obey the normalization
condition 
\begin{equation}
\int_{-\infty }^{\infty }R(x)dx=1.  \label{normalization}
\end{equation}
Thus we exclude asymmetric effects, such as those generated by asymmetric
temporal response functions (with $x$ being time), as in the case of the
Raman effect on optical pulses \cite{Wyller}.

In nonlinear optics (\ref{generalNLS}) represents a general phenomenological
model for media in which the nonlinear refractive index change (or
polarization) induced by an optical beam is determined by some kind of a
transport process. It may include, e.g., heat conduction in materials with a
thermal nonlinearity \cite{thermal1,thermal2,thermal3} or diffusion of
molecules or atoms accompanying nonlinear light propagation in atomic
vapours \cite{suter}. Nonlocality also accompanies the propagation of waves
in plasma \cite{cusp,litvak,df,litvak75,juul}, and a nonlocal response in
the form (\ref{generalNLS}) appears naturally as a result of many body
interaction processes in the description of Bose-Einstein condensates \cite
{bose}.

The width of the response function $R(x)$ relative to the width of the
intensity profile $|\psi|^2(x,z)$ determines
the degree of nonlocality. In the limit of a singular response we get the
well-known nonlinear Schr\"odinger (NLS) equation, which appears in all 
areas of physics. Here the
focusing case ($s$=1) produces MI of the finite bandwidth type,
while the defocusing case ($s$=$-1$) predicts modulational
stability \cite{mi-nlo1}. When the width of the response function is finite
but small compared to the intensity, the model (\ref{generalNLS}) is
approximated by the modified NLS equation 
\cite{KBJW,Parola98,molecular,nakamura,wk-ob} 
\begin{equation}
   i\partial_z\psi + \frac{1}{2}\partial_x^2\psi + s[|\psi|^2+
   \gamma\partial_x^2|\psi|^2]\psi =0.
\label{weaknonlocal}
\end{equation}
Here $\gamma \ll 1$ is defined as the second virial of $R(x)$ as 
\begin{equation}
\gamma \equiv \frac{1}{2}\int_{-\infty }^{\infty }x^{2}R\left( x\right) dx
\label{scaling0}
\end{equation}
and it scales as $\gamma \sim \sigma ^{2},$ where $\sigma $ is the
characteristic length of the response function. In contrast to the local NLS
limit ($\gamma$=0), the MI now depends not only on the sign of $s$, but
also on the intensity of the plane waves \cite{litvak}. 
Finally, in the case of strong nonlocality it has been shown that 
(\ref{generalNLS}) simplifies to a linear model, and hence there 
is no MI in this limit \cite {Snyder97}.

MI has thus been studied in different limits. The general case 
(\ref{generalNLS}) has recently been investigated with respect to MI and 
compared
with the weakly nonlocal limit \cite{KBJW}. Here we present an analytical
study of the full nonlocal case with arbitrary profile $R(x)$ whose spectrum
obeys a sufficient degree of smoothness, with particular emphasis on generic
features of the MI. The present paper complements and extends the results
obtained in \cite{KBJW}.

\section{M$\allowbreak $I in the nonlocal NLS equation}

The model (\ref{generalNLS}) permits plane wave solutions of the form 
\begin{equation}
\psi (x,z)=\sqrt{\rho _{0}}\exp (ik_{0}x-i\omega _{0}z),\quad \rho _{0}>0,
\label{planewave}
\end{equation}
where $\rho _{0}$, $k_{0}$, and $\omega _{0}$ are linked through the
nonlinear dispersion relation 
\begin{equation}
\omega _{0}=\frac{1}{2}k_{0}^{2}-s\rho _{0},  \label{nonlindisp}
\end{equation}
Following \cite{KBJW}, we perturb the plane wave solutions (\ref{planewave})
- (\ref{nonlindisp}) as follows: Assume that 
\begin{eqnarray}
\psi \left( x,z\right) &=&\left[ \sqrt{\rho _{0}}+u\left( \xi ,\tau \right)
+iv\left( \xi ,\tau \right) \right] \exp (i\theta _{0}) \\
\xi &=&x-k_{0}z,\text{ }\tau =z, \\
\theta _{0} &=&k_{0}x-\omega _{0}z,
\end{eqnarray}
where $u$ and $v$ denote the real- and imaginary part of the perturbation.
Inserting this expression into the nonlocal NLS equation (\ref{generalNLS})
and linearizing around the solution (\ref{planewave}) - (\ref{nonlindisp})
yields the equations 
\begin{equation}
\begin{array}{c}
\partial _{\tau }u+\frac{1}{2}\partial _{\xi }^{2}v=0, \\ 
\\ 
\partial _{\tau }v-\frac{1}{2}\partial _{\xi }^{2}u-2s\rho _{0}\int_{-\infty
}^{\infty }R(\xi ^{\prime }-\xi )u(\xi ^{\prime },\tau )d\xi ^{\prime }=0,
\end{array}
\end{equation}
for the perturbations $u$ and $v.$ \ By introducing the Fourier transforms 
\begin{eqnarray}
\widehat{u}(k,\tau ) &=&\int_{-\infty }^{\infty }u(\xi ,\tau )\exp [ik\xi
]d\xi , \\
\widehat{v}(k,\tau ) &=&\int_{-\infty }^{\infty }v(\xi ,\tau )\exp [ik\xi
]d\xi , \\
\widehat{R}(k) &=&\int_{-\infty }^{\infty }R(\xi )\exp [ik\xi ]d\xi ,
\end{eqnarray}
and exploiting the convolution theorem for Fourier transforms, the
linearized system is converted to a set of ordinary differential equations
in $k$-space 
\begin{equation}
\begin{array}{c}
\partial _{\tau }\widehat{u}-\frac{1}{2}k^{2}\widehat{v}=0, \\ 
\\ 
\partial _{\tau }\widehat{v}+\frac{1}{2}k^{2}\widehat{u}-2s\rho _{0}\widehat{%
R}\,\widehat{u}=0,
\end{array}
\end{equation}
which can be written in the compact matrix form 
\begin{equation}
\partial _{\tau }\underline{X}=\underline{\underline{A}}\;\underline{X},
\end{equation}
where the vector $\underline{X}$ and matrix $\underline{\underline{A}}$ are
defined as 
\begin{equation}
\underline{X}=\left[ 
\begin{array}{c}
\widehat{u} \\ 
\widehat{v}
\end{array}
\right] ,\;\underline{\underline{A}}=\left[ 
\begin{array}{cc}
0 & \frac{1}{2}k^{2} \\ 
-\frac{1}{2}k^{2}+2s\rho _{0}\widehat{R}(k) & 0
\end{array}
\right] .
\end{equation}
The eigenvalues $\lambda $ of the matrix $\underline{\underline{A}}$ are
given by 
\begin{equation}
\lambda ^{2}=k^{2}\rho _{0}\left[ s\widehat{R}(k)-\frac{1}{4\rho _{0}}k^{2}%
\right] .  \label{generaleigenvalue}
\end{equation}
The general dispersion relation (\ref{generaleigenvalue}) constitutes the
basis of our study of MI.

Let us summarize the properties of the spectrum $\widehat{R}(k)$:
\begin{enumerate}
\item  
Since $R(x)$ is realvalued and symmetric, then so is $\widehat{R}(k)$, 
i.e. $\widehat{R}(k)=\widehat{R}(-k)=\widehat{R}^{\ast}(k)$.

\item  
The normalization (\ref{normalization}) implies $\widehat{R}(0)$=1.

\item  
The symmetry condition for the Fourier transforms imposes $\widehat{R}'(0)$=0,
i.e. the spectrum has a critical point at $k$=0. Here and in the following 
prime denotes differentiation with respect to the argument.

\item  
The normalization (\ref{normalization}) for $R(x)$ means by assumption that 
$R(x)$ is absolute integrable, and hence by Riemann - Lebesque lemma we have 
$\lim_{k\rightarrow\infty}\widehat{R}(k)=0$ \cite{Folland}.

\item  
The functions $R(x)$, $xR(x)$ and $x^2R(x)$ are assumed to be absolute 
integrable, and thus by \cite{Folland} $\widehat{R}(k)$, $\widehat{R}'(k)$,
and $\widehat{R}''(k)$ are continuous for all $k$.

\item  
The response functions are characterized by typical widths and scaling 
lengths $\sigma$ and they assume the generic form $R(x)=\sigma^{-1}\Phi
(x/\sigma)$, where $\Phi$ is a non-dimensional scaling function.
\end{enumerate}

The spectrum $\widehat{R}\left( k\right) $ can be expressed in terms of the
Fourier - transform $\widehat{\Phi }$ of the scaling function $\Phi $ as 
\begin{equation}
\widehat{R}\left( k\right) =\widehat{\Phi }\left( \sigma k\right) \equiv
\int_{-\infty }^{\infty }\Phi \left( \zeta \right) \exp \left[ i\sigma
k\zeta \right] d\zeta .  \label{scaling1}
\end{equation}
Notice that the list of general properties 1-6 of the spectrum 
$\widehat{R}$ carries over to $\widehat{\Phi}$.

The dispersion relation (\ref{generaleigenvalue}) is now conveniently
rewritten as 
\begin{equation}
\lambda ^{2}/\lambda _{0}^{2}=k^{2}\phi \left( k;s,\sigma \right) ,\text{ }
\label{generaldisp1}
\end{equation}
by means of\ (\ref{scaling1}), the definition 
\begin{equation}
\lambda _{0}\equiv 2\rho _{0}  \label{scaling2}
\end{equation}
the redefinitions 
\begin{equation}
2\sqrt{\rho _{0}}\sigma \rightarrow \text{ }\sigma ,\text{ }\frac{1}{2\sqrt{%
\rho _{0}}}k\rightarrow k  \label{scaling3}
\end{equation}
and 
\begin{equation}
\phi \left( k;s,\sigma \right) \equiv s\widehat{\Phi }\left( \sigma k\right)
-k^{2}  \label{phi}
\end{equation}
The parameter $\sigma $ which measures the width of the response function,
plays the role as a {\em control parameter}. Observe that $\widehat{\Phi }
$ is a continuous differentiable function of $k.$ The crucial point in the
stability analysis is the properties of the function $\phi .$ More
precisely, by appealing to the list of properties 1-6 we can
characterize the set $\Omega $ of $k$ fulfilling the inequality $\phi \left(
k;s,\sigma \right) \geq 0$ for a given value of $\sigma $ as follows:

\begin{itemize}
\item  If $\phi \left( k\right) <0$ for all $k\geq 0,$ then $\Omega $ is
empty. In this case we have modulational stability.

\item  If the {\em transversality condition }$\phi ^{\prime }\neq 0$%
{\em is satisfied at the zeros of }$\phi ,$ the number of zeros is
finite. In addition, these points are distinct and isolated. In this case $%
\Omega $ is given as a {\em finite} {\em union of well-separated,
closed, bounded subintervals} of the positive $k$ - axis. In this case we
have MI of the finite bandwidth type.

\item  The breakdown of the transversality condition for certain values of $%
\sigma $, i.e., 
\begin{equation}
\phi \left( k\right) =\phi ^{\prime }\left( k\right) =0\text{ for }%
k=k_{c},\sigma =\sigma _{c}  \label{bifurcational condition}
\end{equation}
describes bifurcation phenomena like {\em excitation, vanishing, coalescence, 
and separation} of MI bands. A careful
analysis of the behavior of $\phi $ in the neighborhood of the bifurcation
point will reveal the type of phenomena which takes place. The number of
zeros of $\phi $ will change as $\sigma $ passes the critical value $\sigma
_{c}$. The case $\phi ^{\prime \prime }\left( k_{c}\right) >0$ for $\sigma
=\sigma _{c}$, can be interpreted as a condition for two MI bands to merge
together or that one single MI band separates into two gain bands, while $%
\phi ^{\prime \prime }\left( k_{c}\right) <0$ for $\sigma =\sigma _{c}$
represents excitation or vanishing of an MI band. The critical value $\sigma
_{c}$ and the type of bifurcation can be determined as follows: \ Solve the
equation 
\begin{equation}
z\widehat{\Phi }^{\prime }\left( z\right) =2\widehat{\Phi }\left( z\right) ,%
\text{ }z>0  \label{bifurcational condition1}
\end{equation}
Then the widths $\sigma _{c}$ can be expressed as 
\begin{equation}
\sigma _{c}=\frac{z}{\sqrt{s\widehat{\Phi }\left( z\right) }}
\label{critical value}
\end{equation}
provided $s\widehat{\Phi }\left( z\right) >0.$ The second derivative of $%
\phi $ evaluated at the bifurcation point can also be expressed in terms of $%
z:$%
\begin{equation}
\phi ^{\prime \prime }=\frac{z^{2}\widehat{\Phi }^{\prime \prime }\left(
z\right) }{\widehat{\Phi }\left( z\right) }-2  \label{second derivative}
\end{equation}
\end{itemize}

From now on we will assume that the transversality condition is satisfied at
the zeros of $\phi ,$ except for a set of isolated bifurcation points. This
means that the behavior of the graph of $\ $spectrum $\widehat{\Phi }\left(
\sigma k\right) $ relative to the parabola $k^{2}$ determines the stability
properties. We will discuss this aspect in more detail in the coming
subsections. Let us first assume that $\phi \left( k;s,\sigma \right) \geq 0$
for some positive $k$. For such values of $k$ the normalized growth rate $%
\Gamma $ reads 
\begin{equation}
\Gamma =\left| {\cal Re}\left( \lambda \right) \right| /\lambda _{0}=\left|
k\right| \sqrt{\phi \left( k;s,\sigma \right) }
\label{normalized growth rate}
\end{equation}
In the present paper we will consider situations where the MI of the
nonlocal NLS equation is of finite bandwidth type and that each gain band
has a unique maximum growth rate, in the same as we have for the
MI of the focusing NLS model and the weakly nonlocal limit, and it emerges
as a generic property for a large class of response functions like the
Gaussian, the exponentially decay function and the square pulse function. It
is straightforward to derive the conditions which must be fulfilled: We find

\begin{equation}
\Gamma ^{\prime }=\frac{1}{2}\phi ^{-1/2}\left[ 2\phi +k\phi ^{\prime }%
\right]  \label{derivative}
\end{equation}
and hence the maximum point $k=$ $k_{\max }$ obeys 
\begin{equation}
\left[ 2\phi +k\phi ^{\prime }\right] _{k=k_{\max }}=0
\label{critical point}
\end{equation}
The second derivative of $\Gamma $ at $k=$ $k_{\max }$ is given as 
\begin{equation}
\Gamma ^{\prime \prime }=\frac{1}{2}\phi ^{-1/2}\left[ 3\phi ^{\prime
}+k\phi ^{\prime \prime }\right] _{k=k_{\max }}
\end{equation}
and in order to have a local maximum at $k=$ $k_{\max }$ the function $\phi $
must obey the inequality 
\begin{equation}
\left[ 3\phi ^{\prime }+k\phi ^{\prime \prime }\right] _{k=k_{\max }}<0
\label{generic condition}
\end{equation}
The condition (\ref{generic condition}) is referred to as {\em the
generic condition for the existence of a maximum growth rate}. One can prove
that if all the critical points of $\Gamma $ obey (\ref{generic condition}),
these points coincide, and hence we have an MI pattern which resembles the
MI of the local, focusing NLS.

\subsection{The focusing case (s=1)}

In this case the function $\phi $ defined by (\ref{phi}) appearing in (\ref
{generaldisp1}) and (\ref{normalized growth rate}) is given as 
\begin{equation}
\phi \equiv \widehat{\Phi }\left( \sigma k\right) -k^{2}  \label{focusingphi}
\end{equation}
From the list of general properties we find that $\phi \left( 0\right) =1$
and we can prove that the number of zeros of $\phi $, denoted by $N$, is 
{\em odd}, i.e. $N=2m-1$ $\left( m=1,2,..\right) $ and that there are $m$
MI bands in this case. Notice that it always exists a closed, bounded
interval $\left[ 0,k_{1}\right] $ for which $\phi \left( k_{1}\right) =0$
and $\phi \left( k\right) >0$ for $0\leq k<k_{1}.$ We refer to $\left[
0,k_{1}\right] $ as the fundamental gain \ band. This means that we always
have MI in the focusing case, due to the existence of the fundamental band$.$
In all the gain bands the growth rate is given by (\ref{normalized growth
rate}).

Now, assume $\widehat{\Phi }$ is a monotonically decreasing \ function for
all $k\geq 0.$ In this case there is a unique $k_{1}>0$ such that $\phi
\left( k_{1}\right) =0$, $\phi \left( k\right) >0$ $\left( \phi \left(
k\right) <0\right) $ for $0\leq k<k_{1}\left( k>k_{1}\right) $ i.e. there is
only one band producing MI. Now, since $\phi \left( 0\right) =1,$ $\phi
^{\prime }\left( 0\right) =0$ and $\phi \left( k_{1}\right) =0,$ we find
that $\Gamma ^{\prime }$given by (\ref{derivative}) satisfies $\Gamma
^{\prime }\left( 0\right) =1$ and $\lim_{k\rightarrow k_{1}^{-}}\Gamma
^{\prime }\left( k\right) =-\infty .$ Hence, since $\phi $ and $\phi
^{\prime }$ are continuous, we can appeal to the intermediate value theorem
and conclude that there is at least one $k_{cr}$ such that $\Gamma ^{\prime
}\left( k_{cr}\right) =0.$

It is also easy to study the variation of the growth rate curve with the
width $\sigma .$ Simple computation yields 
\begin{equation}
\partial _{\sigma }\Gamma =\frac{1}{2}k^{2}\phi ^{-1/2}\widehat{\Phi }%
^{\prime }\left( \sigma k\right)  \label{width of growth rate}
\end{equation}
and since $\widehat{\Phi }$ is a monotonically decreasing \ function, it
follows that the growth rate decreases with the width $\sigma $ for a fixed
modulation wavenumber $k$. Moreover, the end point $k_{1}$ of the MI band is
a function of $\sigma ,$ and the implicit function theorem yields the change
of $k_{1}$ with $\sigma :$ 
\begin{equation}
\frac{dk_{1}}{d\sigma }=-\frac{\widehat{\Phi }^{\prime }\left( \sigma
k_{1}\right) k_{1}}{\widehat{\Phi }^{\prime }\left( \sigma k_{1}\right)
\sigma -2k_{1}}  \label{velocity1}
\end{equation}
Since $\widehat{\Phi}^{\prime}(\sigma k)\leq 0$ , we will
have $\widehat{\Phi}^{\prime}(\sigma k_1)\sigma-2k_1<0$
(which is the transversality condition $\phi^{\prime}\neq 0$ evaluated at
$k=k_1$). Thus $k_1$ is a decreasing function of $\sigma $, which means
that the width of the MI band decreases in size.

Let us use the Gaussian response function 
\begin{equation}
   R(x) = \frac{1}{\sigma}\Phi\left(\frac{x}{\sigma}\right); \;\;\;
   \Phi(\zeta) = \frac{1}{\sqrt{\pi}}\exp(-\zeta^2) 
   \label{gauss}
\end{equation}
as an example. The Fourier - transform is given by 
\begin{equation}
   \widehat{R}(k) = \widehat{\Phi}(\sigma k) =
   \exp\left[-\frac{1}{4}\sigma^2k^2\right] ,  
   \label{gaussf}
\end{equation}
which is a monotonically decreasing function. In Fig.~1 it appears as a
characteristic feature that the MI is being suppressed as the characteristic
widths $\sigma$ of these response functions are increased. 
Moreover, the MI gain band has a unique maximum growth rate. 
Notice that exactly the same characteristic features are found for the 
exponential response \cite{KBJW}. 

\begin{figure}
  \centerline{\psfig{figure=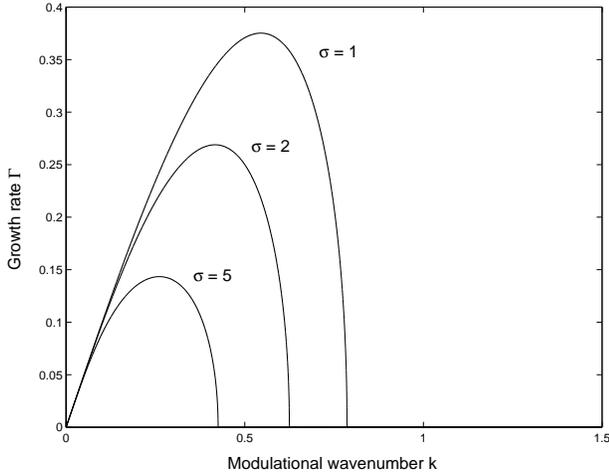,width=\figwidthb}}
  \caption{The growth rate $\Gamma$ as function of the wavenumber $k$ for 
  $\protect\sigma=1,2,5$ and Gaussian response function in the focusing case.}
\end{figure}

Now, let us consider
situations where it is not assumed that the spectrum is strictly decreasing
for all $k$. Then it is possible to have additional gain bands in the
focusing case. We assume that 
\begin{equation}
   \phi \left( k_{2j}\right) =\phi \left( k_{2j+1}\right) =0;  
   \label{a}
\end{equation}
\begin{equation}
   \phi \left( k\right) >0\text{ for }k\in \left\langle
   k_{2j},k_{2j+1}\right\rangle   \label{b}
\end{equation}
for $j=1,2,..,m-1$ where $m$ is the number of MI bands. Since $\Gamma $ $%
\equiv k\phi ^{1/2}$ is continuous and realvalued on $\left[ k_{2j},k_{2j+1}%
\right] $ and differentiable on $\left\langle k_{2j},k_{2j+1}\right\rangle ,$
it follows from Rolles theorem that there is at least one $k_{cr}^{\left(
j\right) }\in \left\langle k_{2j},k_{2j+1}\right\rangle $ such that $\Gamma
^{\prime}(k_{cr}^{(j)})=0$. If, in addition 
\begin{equation}
   \phi^{\prime}(k_{2j}) >0, \;\;\; \phi^{\prime}(k_{2j+1})<0,  
   \label{c}
\end{equation}
we find the following limits
\[
\lim_{\;\;\;\;k\rightarrow k_{2j}^+}  \Gamma^{\prime}(k) = +\infty, \;\;\; 
 \lim_{\;\;\;\;\;\;\;\;k\rightarrow k_{2j+1}^-}\Gamma^{\prime}(k) = -\infty. 
\]

Now, if the condition (\ref{generic condition}) is satisfied, then $%
k_{cr}^{\left( j\right) }=k_{\max }^{\left( j\right) }$ is a unique. Hence
we have a unique maximum growth rate for this wavenumber band as well.

As an example we consider the square pulse function 
\begin{equation}
R\left( x\right) =\frac{1}{\sigma }\Phi \left( \frac{x}{\sigma }\right) ;%
\text{ }\Phi \left( \zeta \right) =\left\{ 
\begin{array}{c}
\frac{1}{2};\text{ }\left| \zeta \right| \leq 1 \\ 
0;\text{ }\left| \zeta \right| >1
\end{array}
\right. ;  \label{square pulse}
\end{equation}
whose Fourier - transform is given by 
\begin{equation}
   \widehat{R}(k) =\widehat{\Phi}(\sigma k) =\frac{\sin(k\sigma)}{k\sigma}.  
   \label{square pulsef}
\end{equation}
For small and moderate values of the width $\sigma$ we only have one 
fundamental MI gain band. 
The bifurcation equation (\ref{bifurcational condition1}) reads 
\begin{equation}
  \tan(z) = z/3  
  \label{bifurcationalsq}
\end{equation}
in this case, and by means of (\ref{critical value}) with $\widehat{\Phi}
(z)>0$ ($\Leftrightarrow \sin(z)>0$), we find the
critical values $\sigma _{c}.$ Moreover, one finds that $\phi^{\prime\prime}
=-z^2-6$, from which it follows that new MI bands are excited at
the bifurcation points. The table below summarizes the findings for the
lowest order excitations. 
\begin{equation}
\begin{tabular}{|c|c|}
\hline
$\sigma _{c,1}=21.03$ & The second MI band is excited \\ \hline
$\sigma _{c,2}=52.555$ & The third MI band is excited \\ \hline
\end{tabular}
\end{equation}
A simple computation based on (\ref{generic condition}) now reveals that
each MI band has a maximum growth rate. Finally, since $\widehat{\Phi}$ 
of the square response function is a monotonically
decreasing function on an interval containing $[0,k_1]$ for
all values of $\sigma $, we conclude by appealing to (\ref{width of growth
rate}) and (\ref{velocity1}) that an increase in the width $\sigma $
decreases both the fundamental band and the maximum growth rate. 
The expression (\ref{velocity1}) (with $k_1$ replaced with $k_i$, 
$(i=2j,2j+1))$ and (\ref{a})-(\ref{c}) show that the higher order MI gain
bands move towards lower wavenumbers as the width $\sigma$ increases. 
Fig.~2 summarizes these features graphically, and the results are consistent
with the findings in \cite{KBJW}.

\begin{figure}
  \centerline{\psfig{figure=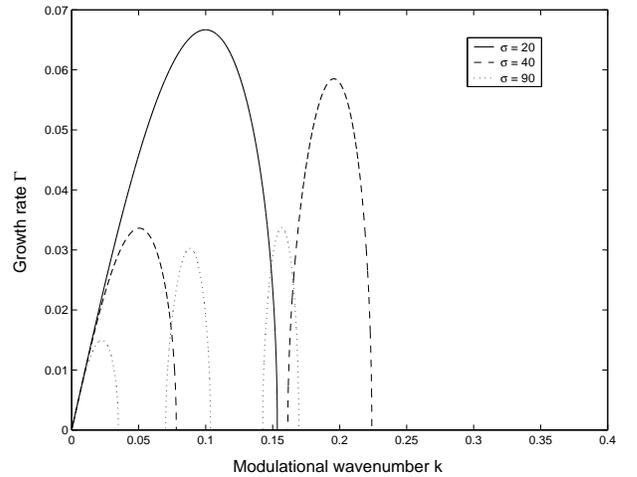,width=\figwidthb}}
  \caption{The growth rate $\Gamma$ as function of the wavenumber $k$ for 
  $\protect\sigma =20,40,90$ and the square response function in the 
  focusing case.}
\end{figure}

\subsection{The defocusing case (s=-1)}

The function $\phi $ given by 
\begin{equation}
\phi \equiv -\widehat{\Phi }\left( \sigma k\right) -k^{2};
\label{defocusingphi}
\end{equation}
in the defocusing case. Here the number of zeros of $\phi $, denoted by $N$,
is {\em even}, i.e. $N=2m$ $\left( m=0,1,2,..\right) $ and there are $m$
MI bands. We observe from the list of properties of the spectrum $\widehat{%
\Phi }\left( \sigma k\right) $ that MI can only exist in the high wavenumber
regime, as opposed to the focusing case. Notice that the case $m=0$
corresponds to the situation where $\phi \left( k\right) <0$ for all $k\geq
0 $ which means modulational stability.

A typical situation revealing stability occurs when $\widehat{\Phi}(
\sigma k)\geq 0$ for all $k\geq 0$, irrespective of the value of $\sigma$. 
The spectrum of the Gaussian (\ref{gaussf}) satisfies this property. 
The same holds true for the exponential response function 
\begin{equation}
   R(x) = \frac{1}{\sigma}\Phi\left(\frac{x}{\sigma}\right); \;\;\;
   \Phi(\zeta) = \frac{1}{2}\exp(-|\zeta|) 
\end{equation}
for which the Fourier - transform is given by 
\begin{equation}
   \widehat{R}(k) = \widehat{\Phi}(\sigma k) = \frac{1}{1+\sigma^2k^2}.
\end{equation}

When $\widehat{\Phi }(k)$ changes sign, the picture
is somewhat more complicated. First of all, we will still have stability if
$\widehat{\Phi}(\sigma k)<0$ for all $k$ provided the control
parameter $\sigma $ is below certain threshold $\sigma_c$. The threshold
value $\sigma_c$ and the corresponding value of $k$ can be found by
requiring the bifurcation condition (\ref{bifurcational condition}) to be
fulfilled. Secondly, if $\sigma $ belongs to the complementary regime, i.e., 
$\sigma>\sigma_c$ we have $\widehat{\Phi}(\sigma k) \geq 0$ for certain 
wavenumber intervals, and hence we may have one or more MI gain bands, whose 
growth rate $\Gamma$ is given by (\ref{normalized growth rate}) and 
(\ref{defocusingphi}). 
Even in this case one can prove the existence of a unique maximum growth 
rate within each unstable band in the same way as in the focusing case.

Again we use the square pulse function (\ref{square pulse}) - (\ref{square
pulsef}) as an example. The bifurcation problem is solved by means of (\ref
{bifurcationalsq}) - (\ref{critical value}) with $s=-1$ and hence the extra
constraint that $\sin \left( z\right) <0.$ Here we also find that $\phi
^{\prime \prime }=-z^{2}-6<0$, which means that new MI bands are excited
through the bifurcation process. The table below displays the bifurcation
values of $\sigma $ for the lowest order excitations.

\begin{equation}
\begin{tabular}{|c|c|}
\hline
$\sigma _{c,1}=9.146$ & The f$\text{irst MI band is excited}$ \\ \hline
$\sigma _{c,2}=35.78$ & The s$\text{econd MI band is excited}$ \\ \hline
$\sigma _{c,3}=71.31$ & $\text{The third MI band isexcited}$ \\ \hline
\end{tabular}
\end{equation}

The number of MI bands increases when the width increases. By (\ref{generic
condition}) each MI band has a maximum growth rate. Finally, by
using the expression for the velocity of the zeros $k_{2j-1}$ and $k_{2j}$
of $\phi $ $(j=1,2,..,m)$ in the defocusing case 
\begin{equation}
\frac{dk_{i}}{d\sigma }=\frac{\widehat{\Phi }^{\prime }\left( \sigma
k_{i}\right) k_{i}}{-\widehat{\Phi }^{\prime }\left( \sigma k_{i}\right)
\sigma -2k_{i}};\text{ }i=2j-1,2j
\end{equation}
and (\ref{a}) - (\ref{c}) with $\phi $ given as (\ref{defocusingphi}) and $%
k_{2j}$ and $k_{2j+1\text{ }}$replaced with $k_{2j-1}$and $k_{2j},$%
respectively, it can be shown in a the same way as in the focusing case that 
$\frac{dk_{i}}{d\sigma }<0$ and hence the MI bands move to the low
wavenumber regimes as the width increases. The latter analytical result is
also in accordance with the numerical results obtained in \cite{KBJW}. In
Fig.3.\ the MI result for the square pulse (\ref{square pulse}) in the
defocusing case is displayed.

\begin{figure}
  \centerline{\psfig{figure=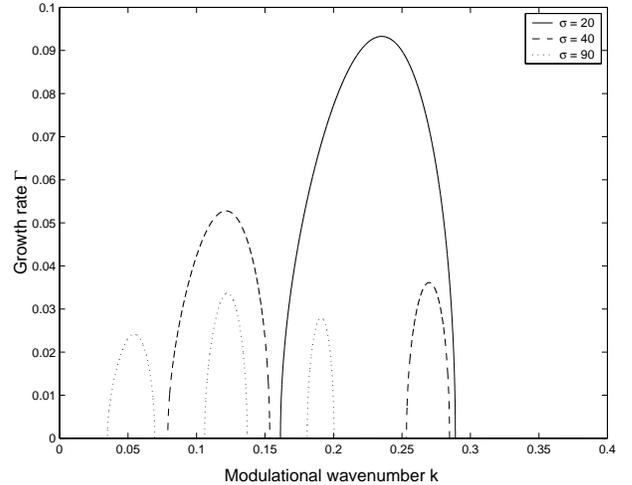,width=\figwidthb}}
  \caption{The growth rate $\Gamma$ as function of the wavenumber $k$ for 
  $\protect\sigma =20,40,90$ for the square response function in the 
  defocusing case.}
\end{figure}

\section{Generalized response functions}

Now, let us consider a localized response function $R(x)$ which can be
written as a convex combination of $N$ response functions $R_{n}\left(
x\right) $ whose properties are known i.e. the weighted mean 
\begin{equation}
R(x)=\sum_{n=1}^{M}f_{n}R_{n}\left( x\right) \text{ , }n=1,2,3,...,M;
\end{equation}
where 
\begin{equation}
\sum_{n=1}^{M}f_{n}=1\text{ , \ }0\leq f_{n}\leq 1\text{ , }\int_{-\infty
}^{\infty }R_{n}\left( x\right) dx=1;
\end{equation}
and $R_{n}\left( x\right) $ is parametrized by a typical lengthscale $\sigma
_{n}$, i.e. $R_{n}\left( x\right) =\sigma _{n}^{-1}\Phi _{n}\left( \frac{x}{%
\sigma _{n}}\right) $ with $\Phi _{n}$ as the scaling functions. From now on
we refer to the functions $R_{n}\left( x\right) $ as ''building blocks''.

The corresponding spectrum $\widehat{R}\left( k\right) $ appears as a convex
combination of $\widehat{R}_{n}\left( k\right) $

\begin{equation}
\widehat{R}\left( k\right) =\sum_{n=1}^{M}f_{n}\widehat{R}_{n}\left(
k\right) =\sum_{n=1}^{M}f_{n}\widehat{\Phi }_{n}\left( \sigma _{n}k\right) ;
\end{equation}
where $\widehat{\Phi }_{n}$ is the spectrum of the scaling function $\Phi
_{n}.$ Moreover, we find in accordance with (\ref{scaling0}) that 
\begin{equation}
\int_{-\infty }^{\infty }x^{2}R_{n}\left( x\right) dx\sim \sigma _{n}^{2};
\end{equation}
A typical lengthscale $\sigma $ of the composite response function $R\left(
x\right) $ can be defined 
\begin{equation}
\sigma ^{2}=\sum_{n=1}^{N}f_{n}\sigma _{n}^{2};  \label{convex1}
\end{equation}
It is straightforward to show that the dispersion relation on normalized
form can be expressed as 
\begin{equation}
\lambda ^{2}/\lambda _{0}^{2}=k^{2}\phi _{conv};\text{ }
\label{convex disprelation1}
\end{equation}
where 
\begin{equation}
\phi _{conv}\equiv s\sum_{n=1}^{M}f_{n}\widehat{\Phi }_{n}\left( \sigma
_{n}k\right) -k^{2};  \label{convex disprelation2}
\end{equation}
which is a generalization of the dispersion relation (\ref{generaldisp1}).
Here $\lambda _{0}$ and $k$ are given as (\ref{scaling2}) and (\ref{scaling3}%
), respectively, while $\sigma _{n}$ is the redefined width of the response
function: 
\begin{equation}
2\sqrt{\rho _{0}}\sigma _{n}\rightarrow \sigma _{n};
\end{equation}
Now, if the transversality condition $\phi _{conv}^{\prime }\neq 0$ is
satisfied on points where \ $\phi _{conv}=0,$ there are a finite union of
disjoint, bounded and closed subintervals of the positive $k$ axis, for
which $\phi _{conv}\geq 0.$ This means that the MI is of finite bandwidth
type in this case as well. In each gain band the normalized growth rate $%
\Gamma $ is given as 
\begin{equation}
\Gamma \equiv \left| {\cal Re}\left( \lambda \right) \right| /\lambda
_{0}=\left| k\right| \phi _{conv}^{1/2};  \label{convex growth rate}
\end{equation}
In (\ref{convex disprelation1}) - (\ref{convex growth rate}) there are $2M-1 
$ {\em control parameters}, namely the $M$ width parameters $\sigma
_{1},\sigma _{2},...,\sigma _{M}$ and the $M-1$ weight parameters $%
f_{1},f_{2},...,f_{M-1}$. To summarize, the stability properties depend on
the weight $f_{n}$ of each Fourier - transform $\widehat{\Phi }_{n}$ and
length scale $\sigma _{n}$ .

Let us consider some special cases. Assume that the Fourier transforms
$\widehat{R}_n(k)$ of all the building block functions $R_n(x)$ are 
monotonically decreasing positive functions, such as the exponential 
and Gaussian response functions. The sum $\sum_{n=1}^{M}f_n\widehat{\Phi}_n
(\sigma_nkt)$ will now be a decreasing and positive function of $k$. 
In the defocusing case $(s=-1)$ we will have modulational stability in 
this case, while we get MI with one gain band about $k=0$ in the focusing 
case ($s$=1), which has a unique maximum growth rate. 
In Fig.~4. we display this phenomenon for the case when the total response
function is a sum of a Gaussian and an exponential response function.

\begin{figure}
  \centerline{\psfig{figure=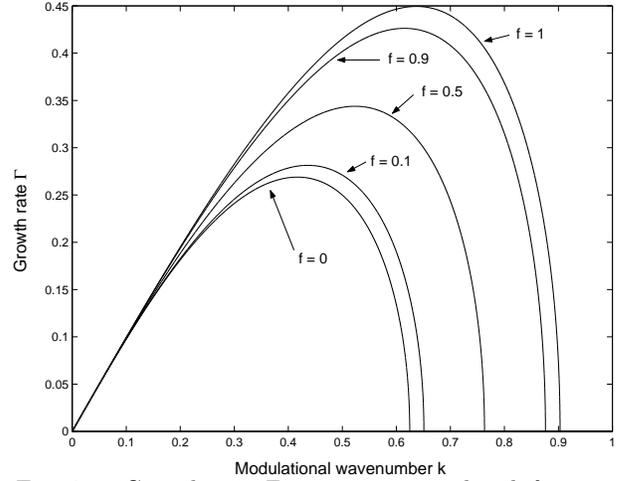,width=\figwidthb}}
  \caption{Growth rate $\Gamma$ versus wavenumber $k$ for a convex 
  combination of a Gaussian response function with weight factor 
  $f_g=f$ and width $\sigma_g$=1 and an exponential response function 
  with weight factor $f_e=1-f$ and width $\sigma_e$=2 for different 
  values of $f$.}
\end{figure}

In Fig.~5 we show what happens for a focusing nonlinearity when the response
function is a 3-component combination of a Gaussian, an exponential, and a 
square response function. In Fig.~5 two MI gain bands are revealed. 
A typical feature of the fundamental band is that the growth rate 
exhibits a more complex behavior with several local extrema, as opposed to
the single-maximum generic situation defined by (\ref{generic condition}).

\begin{figure}
  \centerline{\psfig{figure=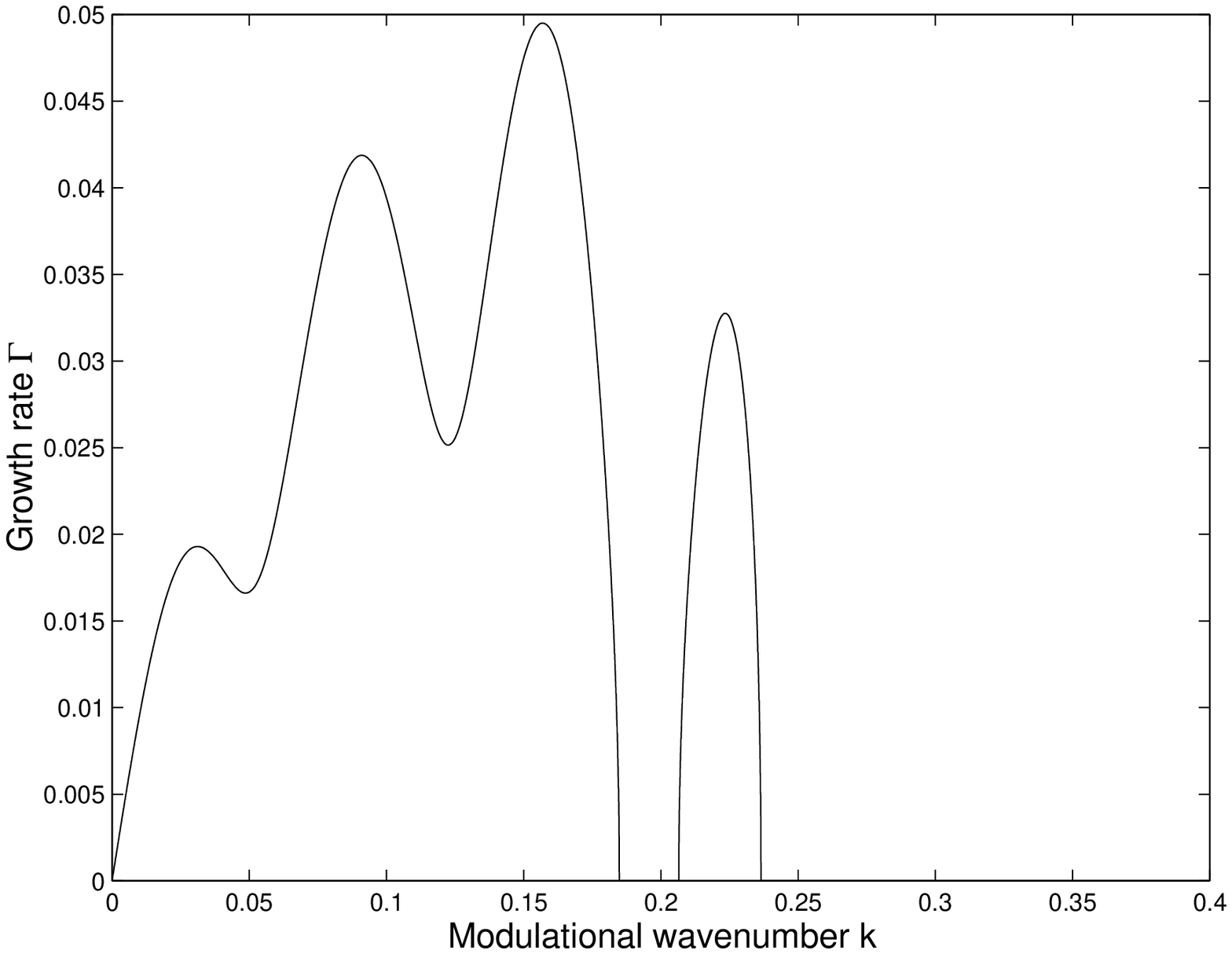,width=\figwidthb}}
  \caption{Growth rate $\Gamma$ versus wavenumber $k$ for a convex 
  combination of a Gaussian (width $\sigma_g$=10), an exponential 
  (width $\sigma_e$=20), and a square response function (width 
  $\sigma_{sq}$=90) in the focusing case. The corresponding weight 
  factors are $f_g$=0.1, $f_e$=0.3, and $f_{sq}$=0.6.}
\end{figure}

\section{Conclusion}

The linear stage of the MI for the nonlocal NLS equation is conveniently
studied in terms of the spectrum of the response function. From the
dispersion relation (\ref{generaleigenvalue}) it appears that the crucial
point in this discussion is the location of the spectrum of the
response function relative to the parabola in $k$-space. The following
features complement and extend the results obtained in \cite{KBJW}:

\begin{itemize}
\item  
The MI is of the finite bandwidth type. It consists of a finite number of 
well-separated gain bands. Moreover, it is possible to predict the
occurrence of excitation, vanishing, coalescence and separation of 
MI bands.

\item  
For a large class of response functions, each MI band has
a unique maximum growth rate. This property holds true for the
Gaussian, the exponential and the square response functions. It
resembles the structure of the MI bands in the focusing local 
NLS equation.

\item  
In the focusing case we always find at least one MI gain band centered 
about $k=0$. It is verified analytically that the width of this MI band, 
as well as the corresponding growth rate, decreases when increasing the
width of the response function, provided the spectrum of the response
function is decreasing in this MI band. 
Furthermore, additional MI bands are excited at higher wavenumbers when 
the width parameter exceeds a certain threshold, i.e. when the nonlinearity
becomes sufficiently nonlocal. 
The latter phenomenon is a unique feature of the nonlocal nonlinearity
and has no equivalent in the local case and the weakly nonlocal limit.

\item  
In the defocusing case we can either have stability or MI of the
finite bandwidth type. The latter situation can only occur in the high
wavenumber regime, and only if the width of the response function exceeds
a certain threshold, i.e. when the nonlinearity becomes sufficiently nonlocal.

\item  
In both the focusing and defocusing case the higher order MI bands
move towards lower wavenumbers as the width of the response function
increases. In the limit of strong nonlocality the MI bands vanish 
completely. This result agrees with the fact the strongly nonlocal 
limit of the NLS model (\ref{generalNLS}) is a linear model.

\item  
It is possible to study MI for more complicated multiscale scenarios
where the response functions decompose into a weighted mean of ''building
blocks'', where each ''building block'' has a characterized width. 
In this case we get a richer repertory of response functions to deal with.
This aims at showing that the results we have obtained represents generic 
features of the MI of the nonlocal NLS with general symmetric,
positive response functions.
\end{itemize}

This work was supported by the Danish Technical Research Council
(STVF - Talent Grant 5600-00-0355), the Danish Natural Sciences
Foundation (SNF - grant 9903273), and the Graduate School in
Nonlinear Science (The Danish Research Academy).

\end{multicols}

\end{document}